\documentclass[reprint,superscriptaddress,showpacs,footinbib,%
amsmath,amssymb,aps,pra,floatfix]{revtex4-1}

\pdfoutput=1      

\usepackage{latexsym,gensymb}
\usepackage{bm}
\usepackage{bbm}  
\usepackage{graphicx}

\usepackage{natbib}   

\usepackage[bookmarks=false,%
colorlinks=true,linkcolor=blue,citecolor=blue,urlcolor=blue]{hyperref}

\begin{document}

\title{Framework for discrete-time quantum walks and
a symmetric walk on a binary tree}
\author{Zlatko Dimcovic}
\email{dimcoviz@onid.orst.edu}
\affiliation{Department of Physics, Oregon State University, Corvallis OR 97331}
\author{Daniel Rockwell}
\affiliation{Department of Mathematics, Oregon State University, Corvallis OR 97331}
\author{Ian Milligan}
\affiliation{Department of Mathematics, Oregon State University, Corvallis OR 97331}
\author{Robert M. Burton}
\affiliation{Department of Mathematics, Oregon State University, Corvallis OR 97331}
\author{Thinh Nguyen}
\affiliation{School of Electrical Engineering and Computer Science, 
Oregon State University, Corvallis OR 97331}
\author{Yevgeniy Kovchegov}
\affiliation{Department of Mathematics, Oregon State University, Corvallis OR 97331}

\date{\today}  

\newcommand{\be}{\begin{equation}}
\newcommand{\ee}{\end{equation}}
\newcommand{\ben}{\begin{equation*}}  
\newcommand{\een}{\end{equation*}}

\newcommand{\ket}[1]{|#1\rangle}
\newcommand{\bra}[1]{\langle#1|}
\newcommand{\projector}[2]{|#1\rangle\langle#2|}
\newcommand{\braket}[1]{|#1\rangle}
\newcommand{\ketbra}[1]{\langle#1|}
\newcommand{\Rto}{\Rightarrow}

\begin{abstract}

We formulate a framework for discrete-time quantum walks, motivated by
classical random walks with memory.  We present a specific
representation of the classical walk with memory $2$ on which this is
based.  The framework has no need for coin spaces, it imposes no 
constraints on the evolution operator other than unitarity, and is 
unifying of other approaches.  As an example we construct a symmetric
discrete-time quantum walk on the semi-infinite binary tree. 
The generating function of the amplitude at the root is computed in
closed-form, as a function of time and the initial level $n$ in the
tree, and we find the asymptotic and a full numerical solution for
the amplitude.  It exhibits a sharp interference peak and a power
law tail, as opposed to the exponentially decaying tail of a broadly
peaked distribution of the classical symmetric random walk on a binary
tree. The probability peak is orders of magnitude larger than it is for 
the classical walk (already at small $n$).  The quantum walk shows
a polynomial algorithmic speedup in $n$ over the classical walk, which
we conjecture to be of the order $2/3$, based on strong trends in data.

\end{abstract}
 

\maketitle


\section{Introduction}

Random walks on graphs (Markov chains) are used extensively in science.
They provide a number of now standard approaches and models in physics.  
Application of such ideas to evolution of quantum systems has led to
the emergence of the field of \textit{quantum walks}, principally
distinguished between discrete-time (DTQW) and continuous-time (CTQW)
quantum walks.  However, being unitary (reversible) processes, quantum
walks are very different from their classical stochastic (Markovian)
counterparts.

Quantum walks are used to approach varied problems, for instance, on
quantum lattice gases, arrow of time, generalized quantum theory,
exciton trapping, or topological phases
\cite{Me96,YCKEK10,MDS05,Metal07,*Th10,KRBD10}. 
They may become a general tool for building physical models; for 
example, see a summary in \cite{KRBD10} and CTQW in transport
phenomena \cite{MB11}.  In quantum computing, they are a universal
primitive \cite{Ch09prl}, and in the algorithmic context, the principle
alternative to quantum Fourier transform. The field has developed since
its initiation \cite{FaGu98,Wa01,ADZ93,ABNVW01,AAKV01}, with established
algorithmic uses, examples of dramatic superiority over classical
approaches \cite{Chea03,ChSV07}, and implementations
\cite{RLBL05,*SMea09,*KFea09,*SCea10,*ZKea10,*BFea10,*Pea10,*Setal11}.
See \cite{Kemp03,Ch09cmp} for a review and a recent summary. 

Standard approaches to quantum walks generally stem from memoryless
classical walks.  However, since quantum evolution is memoried 
(unitary), it seems natural to approach construction of quantum walks
from classical random walks with memory.  Relation between unitarity
and memory in walks has been noted \cite{Me96,MDS05}.
(Also, in computer science memoried and biased approaches are common
and beneficial algorithmically.)

It is our observation that DTQW are most directly related to classical
walks with memory 2.  In this paper we present a general DTQW framework
as a direct analog of a specific representation of memory--$2$
classical walks.  With it we construct a symmetric DTQW on a binary
tree, starting from a pure state at an arbitrary level in the tree,
and compute its amplitude at the root.

Sec. ~\ref{sec-IF} starts with a representation for memoried classical
walks, that is particularly interesting for quantum walks due to the
specific form of the Markov tensor.  In an analogy with it, we then
define a framework for DTQW.  Walks are built by choosing the evolution
operator with no constraints other than unitarity.  They evolve in
the product of state spaces, while the key component of the operator
acts on single states.  There is no need for ``coin'' degrees of
freedom. The framework is flexible and suitable for general graphs.
It is also unifying of other approaches, notably coined and Szegedy's
\cite{Sz04}.

In Sec.~\ref{sec-BT} we apply this framework to a binary tree, a
structure with many uses in physics.  It is a natural environment
for quantum walks, but difficult to utilize with current techniques
for DTQW.  A successful specific construction exists for CTQW
\cite{Chea03}.

We construct a symmetric walk on the semi-infinite binary tree, and
calculate its amplitude at the root, as a function of time and of
the initial level in the tree.  This involves path enumeration
using regeneration structures, manipulated with the $z$--transform.
The obtained closed--form generating function yields the analytic
asymptotic for the amplitude, which we also compute numerically.
The amplitude has a sharp peak and a power law tail, completely
unlike the corresponding classical random walk, and shows a
polynomial speedup in $n$.  The data strongly suggests the order
of the speedup of $2/3$.

A summary is in Sec.~\ref{sec-disc}, and appendices discuss the
steepest descent calculation, and the classical walk.


\section{\label{sec-IF} A framework for discrete-time quantum walks}

The main approach to DTQW follows ideas of classical memoryless walks,
and needs an auxiliary ``coin'' degree of freedom.  Here we employ a
specific representation of memoried Markov chains (\ref{ssec-cRep}),
as an analogy for a general DTQW framework (\ref{ssec-if}) which does
not need coin spaces.  Relation to other approaches is discussed at
the close of this section.  We start with an example that serves as a
motivation for using memoried walks.

\paragraph*{Memoried walks and coined DTQW.}

Classical walks with memory are walks with internal states: apart
from its state, the walk carries other information.  Here we use one
such walk in one dimension, with the following property: the next
step depends on the \textit{direction} of the previous one. If the
walker came to a site $i$ from the site $i-1$, the probability to go
to $i+1$ (to maintain the direction) is $p$, while the probability
to go to $i-1$ (reverse direction) is $1-p$.  This is often called a
\textit{persistent} walk. We now show that coined DTQW are a special
case of a quantum analog of classical persistent walks.

Consider a standard coined walk on a $d$--regular graph with $n$
vertices (for example, \cite{AAKV01}).  The state of the walk is in
the direct product of two Hilbert spaces: an auxiliary (``coin'') space
$\mathcal{H}_{A}$, spanned by $d$ states $\ket{a}$, in which a unitary
operator $C$ mixes components; and a space of vertices,
$\mathcal{H}_{V}$, spanned by $n$ states $\ket{v}$. 
The evolution operator acts in this product space
$\mathcal{H}_{A} \otimes \mathcal{H}_{V}$ as: \
$U  \left(\ket{a}\otimes\ket{v}\right) = S \left(C \otimes I\right) 
\left(\ket{a}\otimes\ket{v}\right)$.  
On a cycle with the $\mathcal{H}_{A}$ basis
$\{\ket{\!\!\uparrow}, \ket{\!\!\downarrow}\}$, 
the shift $S$ can be implemented as \ 
$S = \projector{\!\!\uparrow}{\uparrow\!\!}
\otimes \sum_j \projector{j+1}{j}
\ + \  \projector{\!\!\downarrow}{\downarrow\!\!}
\otimes \sum_j \projector{j-1}{j}
$.
Now consider for $C$ a generalized Hadamard coin operator,
\ben
C = \begin{bmatrix}
\sqrt{p} & \sqrt{1-p} \\
\sqrt{1-p} & -\sqrt{p}
\end{bmatrix},
\een
and evolve the state.  Starting from the pure ``up'' state,
\begin{align} \label{eq-upPers}
& \notag
S\cdot\left(C\otimes I\right) \ket{\!\!\uparrow}\otimes\ket{i} \\
& = \notag
S\cdot \left( \sqrt{p} \ \ket{\!\!\uparrow} \otimes \ket{i} 
+ \sqrt{1-p} \ \ket{\!\!\downarrow} \otimes \ket{i} \right)  \\
& = \sqrt{p} \ \ket{\!\!\uparrow} \otimes \ket{i+1} 
+ \sqrt{1-p} \ \ket{\!\!\downarrow} \otimes \ket{i-1}.
\end{align}
For the pure ``down'' initial state,
\begin{align} \label{eq-downPers}
& \notag
S\cdot\left(C\otimes I\right) \ket{\!\!\downarrow}\otimes\ket{i} \\
& = \sqrt{1-p} \ \ket{\!\!\uparrow} \otimes \ket{i+1} 
- \sqrt{p} \ \ket{\!\!\downarrow} \otimes \ket{i-1}.
\end{align}

\noindent
This is a persistent walk: it maintains the direction with probability 
$p$, and changes it with probability $1-p$. The obtained walk undergoes 
the same spectral analysis as its unbiased special case ($p=1/2$, the 
Hadamard walk).  Walks with a general coin have been studied (for example,
\cite{TFMK03,RSSAAD04,CSL08,XSL09}), as well as persistent (correlated)
walks and their relation to DTQW \cite{MeBl02,Konn09,Stra06}. 
With this example we point out the direct correspondence between them.
Note that the directionality of the walk shows up as soon as the coin 
transformation is allowed to have $p\neq 0.5$.  In other words, the
standard Hadamard transform generally implements a persistent walk
(rather than a memoryless one), only with equal probabilities.

\subsection{\label{ssec-cRep} A representation for classical
memory--2 walks}

Walks with memory $2$ are such Markov processes where the next step
depends on two states: the current one, and the previous one.  Walks
with memory are generally studied by using a suitably enlarged state
space.  In particular, a memory--$2$ Markov chain can be represented
as a memoryless one over the space with $n^2$ states. The transition
matrix is then large ($n^2\times n^2$) and sparse.

Instead, here we represent a Markov chain with memory $k$ by a
probability distribution $\mu(t)$ of dimension $k$, while the Markov
tensor $\mathcal{M}$ is then of dimension $k+1$.  For a memory--$2$
walk over $n$ sites, the space has dimension $n$, and each state is
labeled by two indices (the site the walker came from, and the
current site).  So the probability distribution is two--dimensional, 
\ben
\mu(t) =
\begin{bmatrix}
\mu_{0,0}(t) & \cdots & \mu_{0,n-1}(t) \\
\vdots & & \vdots \\
\mu_{n-1,0}(t) & \cdots & \mu_{n-1,n-1}(t)
\end{bmatrix}.
\een
The matrix $\mu_{ij}$ can also be given by a column of rows $r_i$, or
by a row of columns $c_i$, which we use below.  The following
representation for the third-rank tensor $\mathcal{M}$ and its action
is convenient. 

Let $p_{ij|k}$ be the conditional probability for the transition 
$j\to k$,  given that the walk came to $j$ from $i$.  All transition
probabilities $\{p_{ij|k}\}$ define the evolution operator
$\mathcal{M} = [P_0 P_1\dots P_{n-1}]$, as $n$ layers of $n\times n$
transition matrices $P_j$, $j=0,1,\ldots,n-1$, one for each site:
\ben
P_j = \begin{bmatrix}
p_{0,j|0} & p_{0,j|1} & \cdots & p_{0,j|n-1} \\
\vdots    &           &        & \vdots      \\ 
p_{n-1,j|0} & p_{n-1,j|1} & \cdots & p_{n-1,j|n-1}
\end{bmatrix}.
\een
$P_j$ are by construction transition probability matrices, and
this is the only requirement imposed on them.

The evolution of the state, $\mu_{t+1} = \mu_{t}\, \mathcal{M}$,
with $\mathcal{M}$ acting to the left, is defined as
\ben
\mu(t) \mapsto \mu(t+1) \ \colon \ r_j(t+1) = c_j^{\textsf{T}}(t) P_j,
\een
for each  $j = 0,1,\dots,n-1$, where $r_j$ and $c_j^{\textsf{T}}$
are the $j$-th row and transposed column, respectively, of the matrix
$\mu$.  In words: \textit{
At each site $j$, the $P_j$ associated with that site acts on the
transposed $j$--th column of $\mu(t)$, giving the $j$--th row 
of the evolved $\mu(t+1)$.}

Instead of an $n^2\times n^2$ probability matrix, we use $n$ of
$n\times n$ probability matrices $P_j$.  They implement the evolution:
the $j$--th column of $\mu(t)$ has probabilities to arrive to $j$ from
any site, and after the action of $P_j$ the $j$--th row of $\mu(t+1)$
has probabilities to go from $j$ to any site.  Thus action of all
transition matrices on all columns evolves the probability distribution
over all paths.  The stochastic nature of the process is carried by
the assignment of $\{p_{ij|k}\}$ transition probabilities in $P_j$
matrices 
\footnote{ This representation is not explored in the literature. }.

The $P_j$ transition matrices are simple in most cases of interest.
Consider the cycle, a space $\{0,1,\ldots,n\}$ with identified ends
($0$ and $ n$), with only nearest-neighbor transitions, 
$(j\pm 1,j)\to(j,j\pm 1)$. Take the persistent walk, with probability
$p$ to continue, and $1-p$ to reverse,
\begin{align*}
(j-1,j)\to(j,j+1), & \quad\text{with probability $p$}\\
(j-1,j)\to(j,j-1),  & \quad\text{with probability $1-p$}.
\end{align*}
To obtain this walk, the $P_j$ matrices have the following block
centered at $(j,j)$ ($\mod n$), 
\be  \label{eq-cWalk}
P_j \ = \ \begin{bmatrix}
  \ddots \\
  & 1   \\
  & & 1-p \ & 0 & p   \\
  & & 0   & 1 & 0   \\
  & & p   & 0 & \ 1-p \\
  & &     &   & & 1 \\
  & &     &   & & & \ddots
\end{bmatrix}
\ee
\noindent
The rest of the diagonal of $P_j$ has $1s$, other elements are $0$,
except for the transitions between sites $0\equiv n$, and $n-1$
or 1 (boundary conditions), which are $p_{0,n-1|0}=1-p$ (reverse),
$p_{0,n-1|1}=p$ (continue), etc.  Action of these $P_j$ by the above
prescription carries the walk.


\subsection{\label{ssec-if} Quantum walks: the interchange framework}

The above classical procedure for memory--$2$ walks is directly
elevated to define quantum processes.

Consider a basis in an $N$--dimensional Hilbert space, with vectors
labeled as $\{\ket{i}, i=1,2,\ldots N\}$. They represent states that
the walk is performed on, enumerated on a general graph. The state of
the walk is given in the product $\mathbb{C}^N\times\mathbb{C}^N$
spanned by these bases, by states at the previous ($\ket{i}$)
and current ($\ket{j}$) step,
\be \label{eq-state}
\ket{\psi(t)} = \sum_{ij} c_{ij}(t) \, \ket{i}\otimes\ket{j}.
\ee
The evolution is specified by
\be \label{eq-IF}
\ket{\psi(t+1)} = \, \widehat{U} \widehat{X} \, \ket{\psi(t)},
\quad 
\ket{\psi(t)} = (\,\widehat{U} \widehat{X})^t \, \ket{\psi(0)},
\ee
where $\widehat{X}$ is the interchange operator and $\widehat{U}$ is
defined via unitary operators ${U}_j$ in $\mathbb{C}^N$, assigned for
each site,
\begin{align} \label{eq-Uj_def}
\widehat{X} \colon 
    & \notag \, \ket{i}\otimes\ket{j} \mapsto \ket{j}\otimes\ket{i} \\
\widehat{U} = 
    & \, \sum_{j=0}^{N-1} \, \Pi_j \otimes {U}_j, 
\quad\text{where}\quad \Pi_j = \projector{j}{j}.
\end{align}
$\Pi_j$ selects the first state in the product, and $U_j$ acts on the
second.  Before explicit examples, we make a few general comments.

Consider a pure state of the walk $\ket{i}\otimes\ket{j}$, represented
by an arrow pointing from the previous ($i$) to the current ($j$) site. 
The interchange initiates the walk forward by ``reversing the arrow.''
Then the $U_j$ operator distributes the ``tip of the arrow'' to all sites
the process can access (generally in the subspace of adjacent nodes),
and the evolved superposition is obtained. This is best seen in the
forthcoming example of the binary tree (Fig.~\ref{fig-bt_step}).
The explicit reversal $\widehat{X}$ is crucial; then $U_j$ completely
controls the evolution over site $j$, by acting on the originating
state(s) $\ket{i}$, and sending the process over all paths to a
new state.  The framework does not place any conditions on these
operators, except for unitarity of quantum  evolution. We are free
to choose (or construct) them as needed to implement quantum walks.

Note that this construction needs no mention of classical processes.
The representation of classical memoried walks in Sec.~\ref{ssec-cRep}
is given for motivation and insight, and we now comment on this relation.
The discussed interplay between interchange and (local) $U_j$, critical
for this formulation, has a clear analog in the classical
representation---recall the transposition before (local) $P_j$ evolve
the distribution. Also, the freedom to craft any (unitary) $U_j$ to
implement quantum walks corresponds to a classical property, as
$P_j$ may be any (probabilistic) matrices.  Finally, the classical
representation has no explicit coin toss, and there is no need in
the quantum case to mimic randomization via a coin degree of freedom;
here $U_j$ drive the walk and mix components.


\paragraph*{ Relation to other approaches. }

For a comparison with a memoryless (coined) approach, consider a walk
on the line, over the state space $S=\{\ket{j}, j=0,1,\ldots\}$. 
The quantum walk (\ref{eq-upPers}) and (\ref{eq-downPers})
is obtained with
\ben
U_j(p) = \begin{bmatrix}
 \, \ddots & & & & \\
 \, & { 1} & & & & \\
 \, & & { \sqrt{1-p}} & { 0} 
        & { \sqrt{p}} \\
 \, & & { 0} & { 1} & {  0} &  \\
 \, & & { -\sqrt{p}} & { 0 } & { \sqrt{1-p}} \\
 \, & & & & & { 1} \, \\
 \, & & & & & & \ddots
\end{bmatrix},
\een
with the block centered at $(j,j)$.
The rest of the diagonal has $1s$, and other elements are $0$. (On a cycle
there are elements needed for boundary conditions, like in the classical
case.) The square roots provide for probability being the square of the
amplitude, and the $-\sqrt{p}$ sign is needed for unitarity.  In this 
simple case, the choice of $U_j$ follows from the classical memoried walk
(\ref{eq-cWalk}), but in general a classical analog is not needed.

Now look at the evolution steps from pure states.  Starting from
$\ket{i-1}\otimes\ket{i}$ (the system is in the state $\ket{i}$,
having been in $\ket{i-1}$ at the previous step),
\begin{align*}
\widehat{U}\widehat{X} \, \ket{i-1}\otimes\ket{i} 
& = \, \widehat{U} \, \ket{i}\otimes\ket{i-1} \\
& = \Big( \sum_{j\in S}\projector{j}{j}\otimes U_j \Big)\,
\ket{i}\otimes\ket{i-1} \\
& = \sqrt{1-p} \, \ket{i}\otimes\ket{i-1} 
  + \sqrt{p}   \, \ket{i}\otimes\ket{i+1}.
\end{align*}
Similarly, for the initial $\ket{i+1}\otimes\ket{i}$ state,
\ben
\widehat{U}\widehat{X}\, \ket{i+1}\otimes\ket{i}
= -\sqrt{p} \, \ket{i}\otimes\ket{i-1}+ \sqrt{1-p} \,
\ket{i}\otimes\ket{i+1}.
\een
This is an isomorphism of the memoryless--based walk of
Eqs.~(\ref{eq-upPers}) and (\ref{eq-downPers}), via identification
\ben
\ket{i-1}\otimes\ket{i} \Leftrightarrow 
\ket{\!\!\uparrow}\otimes\ket{i}
\quad\text{and}\quad 
\ket{i+1}\otimes\ket{i} \Leftrightarrow 
\ket{\!\!\downarrow}\otimes\ket{i}.
\een
This analysis applies to arbitrary mixed states, as each component is
evolved separately.  The choice of $p=1/2$ restores the Hadamard walk.
Thus the interchange framework reproduces coined quantum walks on the
line directly, with the above choice for $U_j$.

Memory in quantum walks is mentioned in literature.  For example, it
was noted in study of the classical limit via decoherence and multiple
coins \cite{BCA_pra03,*BCA_pra03_dec}, and a direct relation between
coined walks and classical memoried walks was observed \cite{KBH06}.
Recently a particular ``quantum walk with memory'' \cite{McG10,*KoMa10}
was studied.

An important approach directly resorting to ideas of memoried walks
is the Szegedy walk \cite{Sz04}, which is the most prominent tool in
DTQW not using coin degree of freedom \cite{Ch09cmp}. Its construction
starts from a classical Markov chain, and the resulting evolution 
operator explicitly carries classical transition probabilities. 
It is contained in the interchange framework via the specific choice
\ben
(U_j)_{km} = 2 \sqrt{p(j,i_k) p(j,i_m)} - \delta_{km},
\een
where $p(i,j)$ need to be classical transition probabilities. The present
approach does not require a specific form of the evolution operator.
It is fully defined by (\ref{eq-state})--(\ref{eq-Uj_def}) alone,
without reference to classical walks, and quantum processes with
desired properties are set by choosing $U_j$ without constraints.
Some benefits of this are seen in the next section, where we construct
a symmetric DTQW on a binary tree.  A Szegedy walk on a binary tree
cannot be obtained with equal probabilities for each branch, as there
is no (real) solution for probabilities $p(i,j)$ such as to yield a
$1/3$ probability for the quantum walk.

The Szegedy walk is also a translation of a memoryless walk into a
walk with memory.  The interchange framework is a direct analog of
an explicit representation for memory--$2$ Markov chains.  This is
reflected in some of its properties, discussed above.

The interferometry--motivated \cite{HBF03} scattering walk
\cite{FeHi04} is performed on graph edges, scattering off of vertices
(or subgraphs).  The more general formulation \cite{RHFB09} can
be formally reconciled with  the present framework, while their
designs and interpretations are different, and complementary.
The scattering walk has recently been used, in its form reflecting
the physical (scattering) approach, for certain search problems
\cite{RHFB09,HRB10,Fetal10}, hinting at benefits of coinless
algorithms.  This approach has not been related to stochastic
processes, which underlie the motivation, construction, and expected
uses of the present framework.  We showed here how a very general
formulation of DTQW, unifying of other similar approaches
\footnote{
$U_j$ implements (one-step) memory by acting on the originating
site, it can be interpreted as a scattering operator on $j$, and
it mixes components (\`{a} la coined walks).
},
arises naturally from Markov chains with memory.

Before moving to a full example we remark on the versatility of this
approach. It reproduces coined walks on the line and Szegedy's walk,
and handles a walk on a binary tree (next section)---each by a simple
choice for $U_j$. This demonstrates flexibility, and it seems that
the framework can help with problems that so far have been prohibitively
difficult.


\section{\label{sec-BT}
A symmetric discrete-time quantum walk on a binary tree}

\begin{figure}[!ht]
\includegraphics[width=2.2in,clip=true]{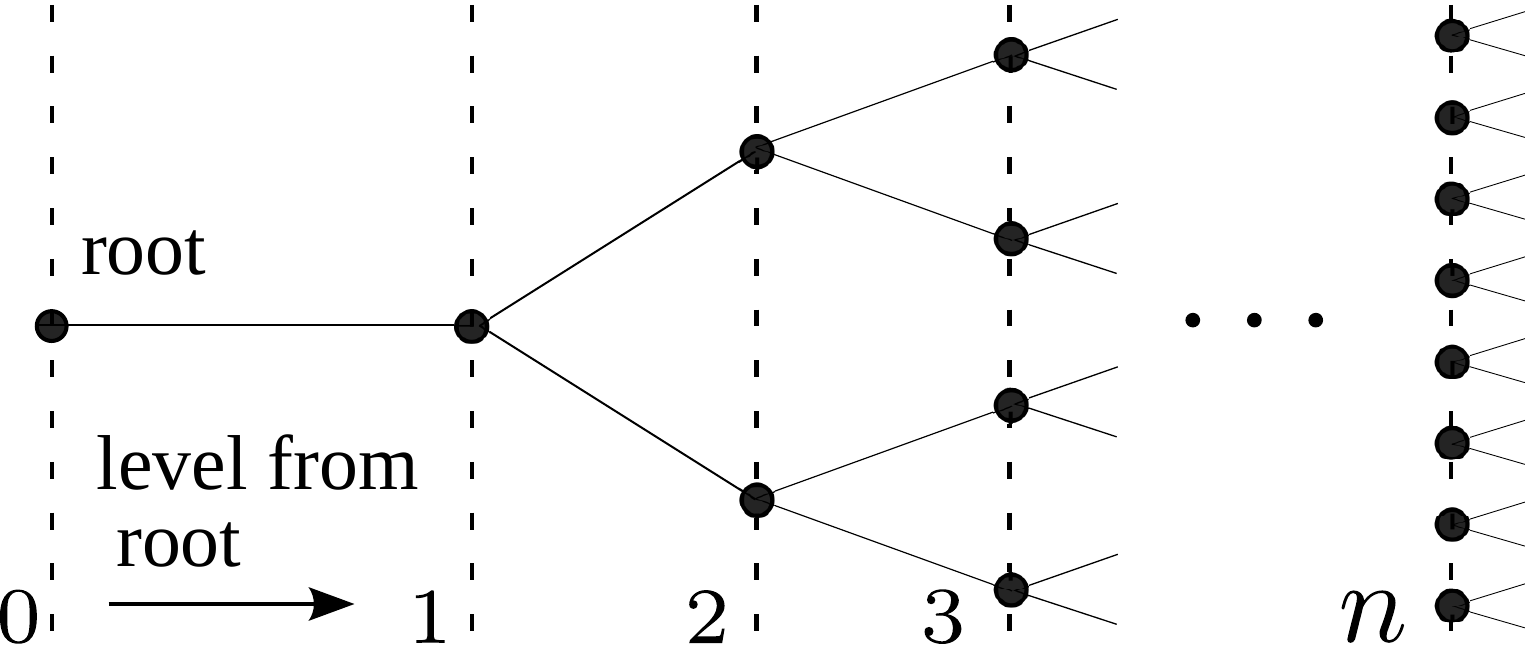}
\caption{ \label{fig-tree} Conventions used for our binary tree.}
\end{figure}

The binary tree is a common model in physics, and a structure of interest
in quantum computing 
\cite{*[{For instance: }] [] FGG08,*SDV06,*Retal09,*Setal10}.  One of
the initiating works \cite{FaGu98} used it as a model for decision
trees, and one of the most successful algorithms \cite{Chea03} solves
a particular problem on connected binary trees, both using CTQW. We
have not seen such progress in using DTQW on the binary tree, even
though this would be beneficial for many problems. This seems to be
due mostly to trouble in handling coin spaces that are necessary for
(coined) DTQW.  In this section we use the established framework to
set and calculate a symmetric DTQW on the semi-infinite binary tree.
We orient our tree with the root (single starting node) at the left
with the tree spanning to the right (Fig.~\ref{fig-tree}).

We focus on the following basic question. The walk is started from a
pure state at a site in the tree at a level $n$, and we examine its
amplitude at the root as a function of time (step) and the initial
distance $n$.

\begin{figure}[!ht]
\includegraphics[width=3.2in,clip=true]{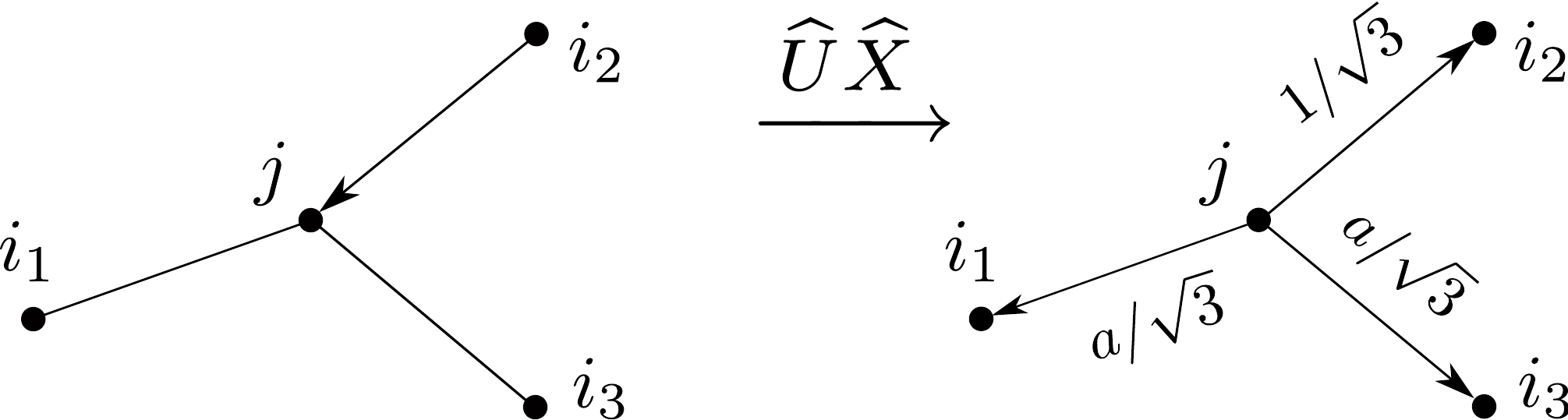}
\caption{   \label{fig-bt_step}
A step taken from a pure state $\braket{i_2}\otimes\braket{j}$
[Eqs.~(\ref{eq-Uj})--(\ref{eq-evolStep})].  The state 
is sent by $\widehat{U}\widehat{X}$ over all available paths.
Probabilities for either branch are chosen to be equal, regardless
of how the walk approaches the site $j$ (the walk is symmetric).
Each component of a general (mixed) state is evolved this way.
}
\end{figure}

The desired symmetric walk has equal probability to step to either of
the connecting nodes, having come from either direction. (The analysis
remains unchanged for different choices of local $U_j$.) The state of
the walk is given in the direct product of spaces, each spanned by
states defined at nodes $S=\{\ket{i}\}$. Label a node in the tree as
$j$, and the nodes connected to it as $i_1$ (to its left, toward the
root), $i_2$, and $i_3$ (to its right, away from the root), as in
Fig.~\ref{fig-bt_step}.  Consider an evolution step from a pure state
at $j$, for example, $\ket{\psi_0}=\ket{i_2}\otimes\ket{j}$. The action
of the interchange $\widehat{X}$ reverses the state. Next we want to
write down the $U_j$ matrix, acting on $\ket{i_2}$, such that the
evolved state has equal probabilities for either branch.  Formally
$U_j$ operate in the space of all nodes, but they are reduced to
the subspace of nodes to which transitions are allowed; here the
adjacent ones.  Thus $U_j$ can be written in a block-diagonal form,
with the non-trivial transition matrix
$U_j^{\,\text{red}}$ in $\{\ket{i_1},\ket{i_2},\ket{i_3}\}$,
and an identity matrix over the remaining dimensions.  The matrix
elements need to satisfy the unitarity of $U_j$ and equal
squared amplitudes of components of the state evolved by it.
The obtained evolution operator, with (reduced) transition matrix
$U_j^{\,\text{red}}$  in the basis $\{\ket{i_1},\ket{i_2},\ket{i_3}\}$,
is
\be \label{eq-Uj}
U_j = \left[ \begin{array}{cc} U_j^{\,\text{red}}
& 0 \\ 0 & \mathbb{I} \end{array} \right],
\quad\text{with} \quad U_j^{\,\text{red}}
= \frac{1}{\sqrt{3}} \begin{bmatrix}
\, 1 & a &  a \, \\
\, a & 1 &  a \, \\
\, a & a &  1 \,
\end{bmatrix},
\ee
where $a = e^{ 2\pi\bm{i}/3 }$.  This representation holds for
graphs of any degree, where dimensions of $U_j^{\,\text{red}}$ and
$\mathbb{I}$ change.  At the root the walk can only get reflected,
which is performed by interchange $\widehat{X}$; then $U_0$ is the
identity matrix.  This will be accounted for.  For all other states,
we now follow the prescription of Eqs. (\ref{eq-IF}) and 
(\ref{eq-Uj_def}). 
With $\widehat{U}=\sum_{i\in S}\Pi_{i} \otimes U_{i}$, the step is
\begin{align}
\ket{\psi_1} = \, & \notag
\widehat{U}\widehat{X}\, \ket{\psi_0} = \left(
\sum_{i\in S} \projector{i}{i}\otimes {U}_{i} \right)
\widehat{X} \, \ket{i_2}\otimes\ket{j} \\ 
= \, & \ket{j} \otimes U_j \ket{i_2} = \,
\ket{j} \otimes \frac{1}{\sqrt{3}} \,
(\,a\,,1\,,a\,,0\,,\hdots)^{\textsf{T}}.
\end{align}
Thus the state is evolved by $\widehat{U}\widehat{X}$ to the
superposition
\be \label{eq-evolStep}
\braket{i_2}\otimes\braket{j} \to \braket{j}\otimes \left(
\frac{a}{\sqrt{3}}\, \braket{i_1} + \frac{1}{\sqrt{3}}\, \braket{i_2}
+ \frac{a}{\sqrt{3}}\, \braket{i_3}\right).
\ee
Each component of the superposition takes the next step from its node
in the same way, and the process spreads over the tree 
\footnote{
For a mixed state at $j$, $U_j$ acts on each originating state,
\( {
\widehat{U}\widehat{X} \, \ket{\psi}^{(j)} =
\widehat{U}\widehat{X}
\sum_{i\in S} c_{ij} \, \ket{i}\!\otimes\!\ket{j}
= \sum_{i\in S} c_{ij} \, \ket{j} \otimes U_j \ket{i}
} \),
fully specifying propagation over site $j$. We only need the evolution
step of a pure state for this calculation. 
}.

In comparison with Markov processes, a quantum walk can be considered
as the evolution of the amplitude distribution.  Also, the causality
typical of the local dynamics of the classical Markov evolution, seen
in Sec.~\ref{ssec-cRep}, is reflected in the quantum walk.  Note how
the concerted action of $\widehat{X}$ and $U_j$ implements the
``arrowed'' (memoried) nature of the evolution, mentioned in 
Sec.~\ref{ssec-if}.

For organizing the calculation, it is useful to note the connection
between directionality and weights of the components of the evolved
state. The component that reverses the direction of the previous step
has the coefficient $1/\sqrt{3}$, while the other two have $a/\sqrt{3}$
(see Fig.~\ref{fig-bt_step}).  This is always the case for this walk
(not only for $\ket{i_2}\otimes\ket{j}$), since it is symmetric,
as explicit in Eqs.~(\ref{eq-Uj})--(\ref{eq-evolStep}).

\paragraph*{Outline of the calculation.}
The amplitude at the root at time $t$ is computed as a sum of the
contributions (amplitudes) of all possible (classical) paths that
are at the root at that time. This is practically a discrete form of
the path integral in quantum mechanics, and is a standard technique
\cite{Me96,ABNVW01}.  So we count all such classical paths on this
structure, weighted appropriately.

The presence of a reflective boundary (the root) complicates the
classification and counting, and we use regeneration structures,
which are then handled via the $z$ transform. The obtained explicit
expression for the transform is complicated, and analytically the
asymptotic of its inverse is found, using the method of steepest
descent. The full amplitude is calculated numerically.

For brevity in involved descriptions, we sometimes use ``paths
$h(t)$" to refer to ``those paths that contribute to the part of
the amplitude (that is named) $h(t)$.''

\subsection{Path counting and regeneration sums}

Enumeration of paths, weighted with appropriate coefficients 
(amplitude), is a combinatorial problem.
Given the symmetry between up and down directions, the tree can be
projected to a line bisecting it.  The paths on the tree can be
classified, and this results in rules for an equivalent walk on 
that line. 

A component of the state at a site is directed either toward or away
from the root; and it can either continue in the same direction or
reverse it in the next step.  For example, the component directed
toward the root (to the left) can continue toward the root (taking
the branch to the left), with the amplitude $a/\sqrt{3}$, or it can
turn and step away from the root, by directly reversing or (and)
by taking the other branch leading away, with the total amplitude
of $(1+a)/\sqrt{3}$ [see Fig.~\ref{fig-bt_step}
or Eq.~(\ref{eq-evolStep})].  

Summarized by the direction of the previous step, the walk on the
line can take the next step as follows:

\noindent
When directed away from the root (to the right), it can:

(i) turn back, with the coefficient $1/\sqrt{3}$ (left turn);

(ii) continue, with $(a+a)/\sqrt{3}$ (right step).

\noindent
When directed toward the root (to the left), it can:

(i) turn away, with $(1+a)/\sqrt{3}$ (right turn);

(ii) continue, with $a/\sqrt{3}$ (left step).

There is a special case, not following the above classification, which
complicates the counting of paths considerably.  We need to count
weighted paths that are at the root at time $t$. Paths generally
reach the root in fewer than $t$ steps, then going back and forth in
the tree, possibly touching the root again in the process, before
finally finding themselves at the root at time $t$.  Whenever they
touch the root their next step can only be a turn back, with the
coefficient $1$, and this does not fall into the above classification.
To account for it the paths need be enumerated particularly carefully.

\begin{figure}[!ht]
\includegraphics[width=3.2in,clip=true]{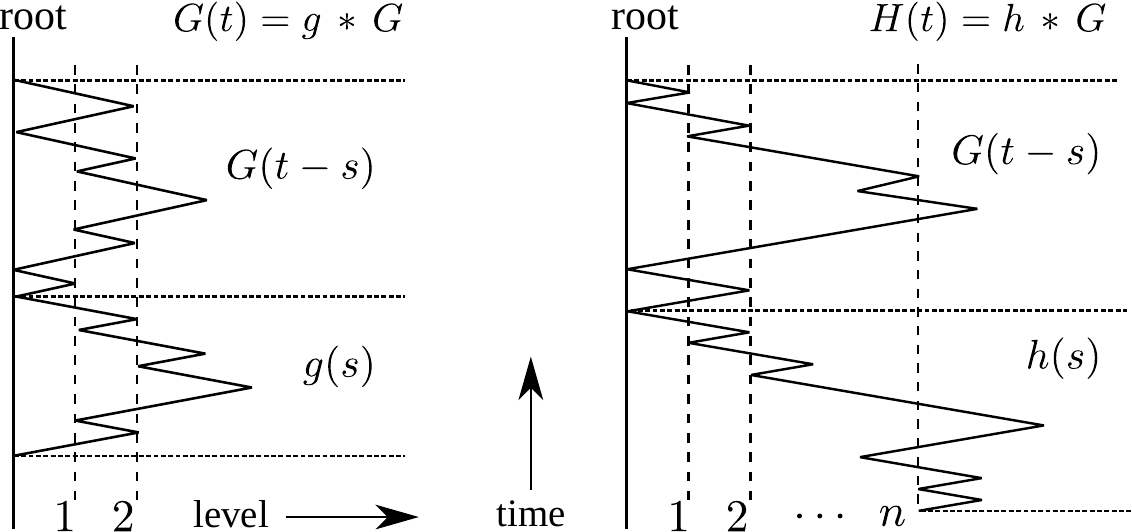}
\caption{  \label{fig-paths}
The number of paths is a convolution  w.r.t. the first contact with
root, with the regeneration structure of Eqs. (\ref{eq-Ht}) and 
(\ref{eq-Gt}).
}
\end{figure}

All paths that are at the root at time $t$ have the following structure.
They touch the root for the first time at one point (step $s$), and we
call the amplitude for this part of the path $h_n(s)$. Then they go out
in the tree, eventually coming back to the root at step $t$, possibly
touching it multiple times in the process; we call the amplitude of this
part of the path $G(t-s)$.  This is encoded by the convolution over the
first contact with the root, and the amplitude, represented by weighted
paths that are at the root at step $t$, starting from level $n$, is
\be \label{eq-Ht}
H_n(t) = \sum_{s\geq n}^{t} h_n(s) G(t-s).
\ee
After the root is touched for the first time, the remainder of the walk 
is a root--to--root path, considered independently as $G(t)$ [accounting
for the $n=0$ case, $G(t)=H_0(t)$].  It consists of: a ``simple loop''
$g(s)$, that goes from the root into the tree and back to it (reaching
it again for the first time), followed by the rest of the path $G(t-s)$,
which may touch the root multiple times, so again comprised of simple
root--to--root loops (Fig.~\ref{fig-paths}).

\noindent
This is a convolution too, with one adjustment.  To correctly account
for $t=0$, it must be that $g(0)=0$, as well as $G(0) = 1$ (since
$t=0\Rto s=t$).  Thus $\delta_0(t)$ need be added. The amplitude of
such root--to--root multi--loops $G(t)$ is
\be \label{eq-Gt}
G(t) = \sum_{s=0}^{t} g(s) G(t-s) + \delta_0(t).
\ee
The regeneration sums, Eqs. (\ref{eq-Ht}) and (\ref{eq-Gt}), organize
path counting.  We work with them using the $z$ transform,
\ben
\widehat{f}(z) = \sum_{t=0}^{\infty} f(t) z^t, \quad |z| < 1.
\een
Applying this transformation to Eqs.~(\ref{eq-Ht}) and (\ref{eq-Gt}),
\begin{align} \label{eq-Hhat}
&  \widehat{H}_n(z) = \, \notag
\widehat{h}_n(z) \, \widehat{G}(z), \ \text{and} \
\widehat{G}(z) = \widehat{g}(z) \, \widehat{G}(z) + 1, \\
& \Rto \quad  
\widehat{H}_n(z) = \widehat{h}_n(z) \, \frac{1}{1-\widehat{g}(z)}.
\end{align}
We need transforms of amplitudes of paths reaching the root for the
first time [$\widehat{h}_n(z)$], and of simple loops [$\widehat{g}(z)$].

At this point we note a known combinatorial result.  The number of paths
on a lattice in two dimensions, going from $(0,0)$ to $(2n,0)$, taking
only Northeast or Southeast steps, with $k$ peaks, is given by Narayana
numbers \cite{Mac15},
\be \label{eq-Narayana}
N(n,k) = \frac{1}{n} \binom{n}{k}\binom{n}{k-1}.
\ee
This expression applies to the number of paths comprising the simple
loops $g$, where peaks are positions furthest from the root. We first
need to identify and enumerate ``steps'' and ``turns''
in such paths, so that we can assign weights to them accordingly.

A simple loop must take an even number of steps.  The first step is a
reversal: it starts at the root, having arrived to it from the first
node, and it can only step back onto the first node, so the coefficient
for this step is $1$.  It is straightforward to establish that paths
with $k$ peaks take $k$ left turns and $k-1$ right turns.  Also, loops
of $t$ steps must take $\frac{t-2}{2}-(k-1)$ right, as well as left,
steps.  Loops with $t=2$ are different: they can only step away from
the root and return to it in the next step (left turn);
their coefficient is $1\times 1/\sqrt{3}$. Thus a simple loop with $k$
peaks, for $t \geq 4$ steps, bears the coefficient:
\begin{align*}
& \left(\frac{1}{\sqrt{3}}\right)^k 
  \left(\frac{1+a}{\sqrt{3}}\right)^{k-1}
  \left(\frac{a}{\sqrt{3}}\right)^{t/2-k}
  \left(\frac{2a}{\sqrt{3}}\right)^{t/2-k} \\
& \quad = 
  \frac{1}{(\sqrt{3})^{t-1}} \left(\frac{1+a}{2a^2}\right)^{k-1}
  \left(2a^2\right)^{t/2-1} \\
& \quad = 
  \, \frac{1}{(\sqrt{3})^{t-1}} \left(\frac{-1}{2}\right)^{k-1}
  \left(2a^2\right)^{t/2-1} \quad\text{(as $a=e^{2\pi i/3}$)}.
\end{align*}
For $a=e^{2\pi\bm{i}/3}$ we have $1+a+a^2 =0$, used above.  Summed
over all possible numbers of peaks $k$, and with the $t=2$ case
added, the amplitude of a simple loop is 
\begin{align*}
g(t) = \ & \frac{1}{\sqrt{3}}\, \delta_0(t-2) \\
         & + \ \sum_{k=1}^{\frac{t-2}{2}} 
\frac{ \left(2a^2\right)^{t/2-1} }{(\sqrt{3})^{t-1}}
\left(\frac{-1}{2}\right)^{k-1} \, N\left(\frac{t-2}{2},k\right).
\end{align*}
Using the Narayana numbers (\ref{eq-Narayana}), with $t=2m+2$,
\begin{align*}
g(m) = \frac{1}{\sqrt{3}}\, \delta_0(m) \ + \ &   
\frac{1}{m\sqrt{3}} \left(\frac{2a^2}{3}\right)^{\!m} \\
& \times 
\sum_{k=0}^{m-1} \left(\frac{-1}{2}\right)^k 
\binom{m}{k+1} \binom{m}{k}.
\end{align*}
Since we will need the transform of $g(t)$, it is helpful to write
the above sum as an integral, using the identity
\begin{align*}
&  
\sum_{k=0}^{m-1} (\alpha\beta)^k \binom{m}{k+1} \binom{m}{k} = \\
& \qquad \frac{1}{2\pi}\int_0^{2\pi} \left(1+\alpha e^{ix}\right)^m
\left(1+\beta e^{-ix}\right)^m \frac{e^{-ix}}{\alpha} dx.
\end{align*}
Employing this, under the constraint $\alpha\beta = -1/2$,
\begin{align} \label{eq-gm}
g(m) = \ & 
\frac{1}{\sqrt{3}}\, \delta_0(m) + \frac{1}{2\pi} \frac{1}{m\sqrt{3}}
\left(\frac{2a^2}{3}\right)^{\!m} \\
& \notag \times \int_0^{2\pi} \left(\frac{1}{2} + \alpha e^{ix} + 
\beta e^{-ix}\right)^m \frac{e^{-ix}}{\alpha} dx.
\end{align}
It is calculationally convenient to take the $z$--transform of
$g(t)$ at this point.  Since loops take even number of steps,
and $\widehat{g}(z)_{t=0} = g(0) = 0$, with $t=2m+2$,
\begin{align*}
\widehat{g}(z) = & \sum_{t=0}^{\infty} g(t) z^t  = g(0) + g(2) z^2 +
\sum_{t=4,6,\ldots} g(t) z^{t} \\
= & \ 
\widehat{g}(z)_{m=0} + \sum_{m=1}^{\infty} g(m) z^{2m+2}.
\end{align*}
The transform of $\delta$ is $1$, and Eq.~(\ref{eq-gm}) becomes
\begin{align*}
\widehat{g}(z) = \, & 
\frac{1}{\sqrt{3}} z^2 + \frac{1}{2\pi\sqrt{3}}\, z^2 \int_{0}^{2\pi}  
dx \, \frac{e^{-ix}}{\alpha} \\
& \times \left[
\sum_{m=1}^{\infty} \frac{1}{m} \left(\frac{2a^2}{3}\right)^{\!m}
\left(\frac{1}{2} + \alpha e^{ix} + \beta e^{-ix}\right)^m z^{2m}
\right].
\end{align*}
Now we make use of $\sum_{n=1}^{\infty} x^n/n = - \ln(1-x)$, $|x|<1$,
and at this point pick $\alpha=-\beta=1/\sqrt{2}$, arriving at
\begin{align*}
\widehat{g}(z) = \frac{z^2}{\sqrt{3}} + \, &
\frac{z^2}{2\pi} \, \sqrt{\frac{2}{3}}  \, 
\int_0^{2\pi} dx \ e^{-ix} \\
& \times \left\{ - \ln\left[ 1 - z^2\frac{2a^2}{3}
\left(\frac{1}{2} + \frac{e^{ix}-e^{-ix}}{\sqrt{2}}\right)\right]
\right\}.
\end{align*}
Using $\omega = e^{-ix}$ and integrating by parts,
\begin{align*}
\widehat{g}(z) = \, \frac{1}{\sqrt{3}} z^2 \ - \ 
& z^4 \, \frac{1}{2\pi i} \, \frac{2a^2}{3} \\
& \times  \oint\limits_{|\omega|=1}
\frac{  \left( \frac{1}{\omega} + \omega \right) d\omega
}{  1 - \frac{2a^2 z^2}{3} \left[\frac{1}{2} + \frac{1}{\sqrt{2}}
\left(\frac{1}{\omega} - \omega\right) \right]
}.
\end{align*}
Here the Residue Theorem is used. The singularity at $\omega=0$ is
removable, while one of the two zeros of the denominator is inside
the integration contour.  Finally, 
\be \label{eq-ghat}
\widehat{g}(z) = \frac{\sqrt{3}}{2a^2} \left[
1 + \frac{1}{3} (az)^2 - \sqrt{1 - \frac{2}{3} (az)^2 + (az)^4} 
\right].
\ee
This closed-form expression analytically extends $\widehat{g}(z)$
beyond the disk $|z|<1$ on which it was defined.  We now need to
deal with $\widehat{h}_n(z)$. 

Paths from the $n$-th level that reach the root for the first time in
$s$ steps, with amplitude $h_n(s)$, first reach the level $n-1$,
generally going out into the tree in the meanwhile, then the level
$n-2$, and so forth until the root is hit.  This is organized into
paths dropping by one level closer to the root (with amplitude $h_1$),
convoluted with the rest of the walk, which itself is comprised of
paths getting closer to the root by one level, 
$h_n = h_1 \ast h_{n-1} = \ldots = h_1 \ast \ldots \ast h_1$
($n$ times). Then the transform is
\be \label{eq-hhat}
\widehat{h}_n(z) = \left[ \widehat{h}_1(z) \right] ^n.
\ee
Paths $h_1$, that for the first time reach one level closer to the
root, are combinatorially equivalent to the paths that start at level
$1$ and touch the root for the first time. 

\begin{figure}[!ht]
\includegraphics[width=3.2in,clip=true]{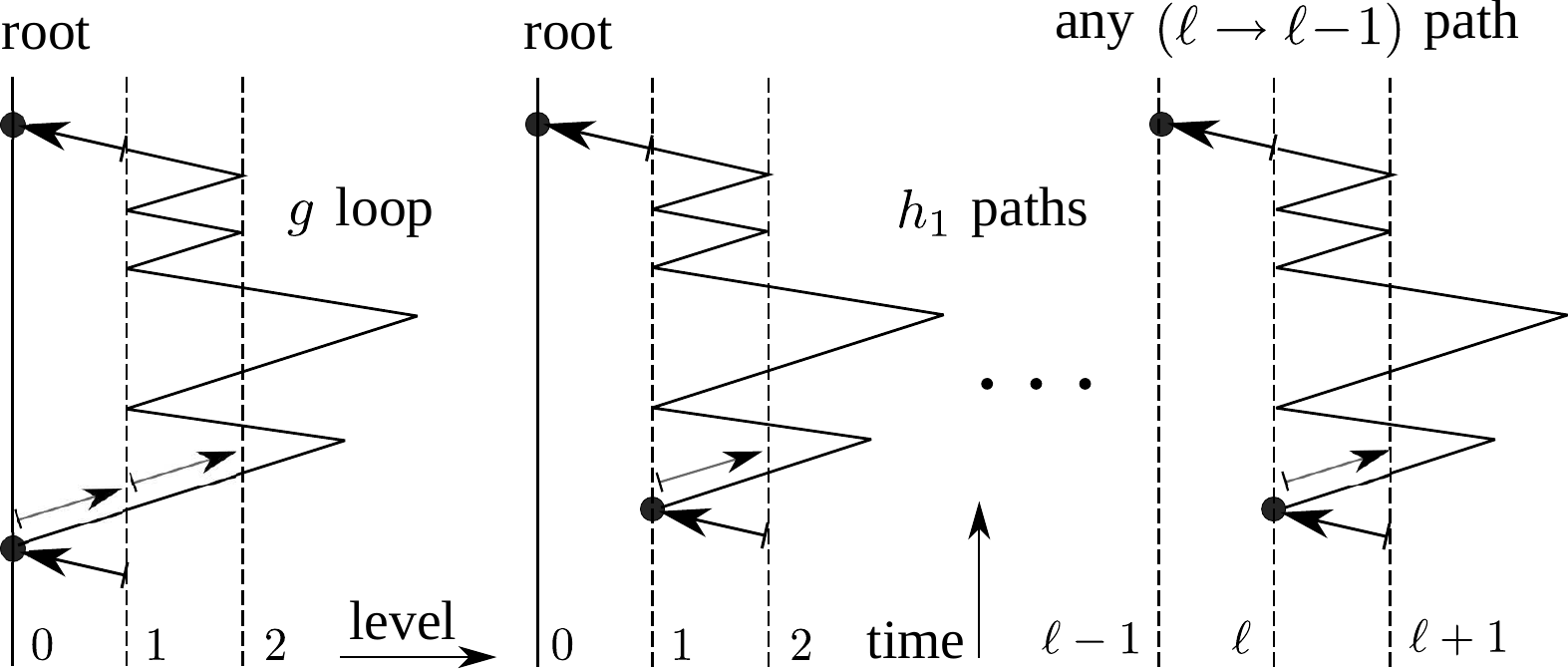}
\caption{\label{fig-h1}
Combinatorial comparison of root--to--root loops ($g$) and paths
getting closer to root by one level ($h_1$) (see text).
}
\end{figure}

Consider such a path, starting at level $1$ and finding its way to
the root for the first time, in more than one step (Fig.~\ref{fig-h1}).
It must have come to level $1$ from level $2$, and it first steps back
to level $2$.  For comparison, now recall a root--to--root simple loop
$g(s)$.  It differs from $h_{1}$ by: the first step of $g$ (with
coefficient $1$) is not taken by $h_{1}$ (which is already at level
$1$), and the next step of $g$ (for paths with $t>2$) is a
right step, while the $h_{1}$ path takes a right turn. 

\noindent
So we divide the expression for $g(s)$ by the coefficient associated
with the second step of $g$ that $h_{1}$ does not take, $2a/\sqrt{3}$,
and multiply it by the coefficient of the step that $h_{1}$ takes
instead, $(1+a)/\sqrt{3}$.  We also divide by the coefficient of the
first step of $g$, not taken at all by $h_{1}$, which is $1$. In the
special case $s=1$ a single step is taken to the root from the first
level, with $a/\sqrt{3}$. Finally, this path takes one step more as
compared to $g(s)$, so we use the expression for $g(s+1)$, starting
from $s=3$ since $g(2)$ corresponds to the special case $h_{1}(1)$.
This gives us the expression for the amplitude $h_1$,
\ben
h_1(s) = \frac{a}{\sqrt{3}} \delta_0(s-1) 
+ \frac{ \frac{1+a}{\sqrt{3}} }{ 1 \times \frac{2a}{\sqrt{3}} }
\,  g(s+1) \times \mathbbm{1}_{s\in\{3,5,\ldots\}} . 
\een
Its transform is, using $1+a+a^2=0$ (as $a=e^{2\pi\bm{i}/3}$),
\begin{align} \label{eq-h1hat}
\widehat{h}_1(z) & = \frac{a}{\sqrt{3}} \, z + \frac{1+a}{2a}
\left[ \frac{1}{z} \sum_{t=3,5,\ldots} z^{t+1} g(t+1) \right] \\
&  \notag = \frac{a\,z}{\sqrt{3}} - \frac{a}{2\,z} 
\left( \widehat{g}(z) - \frac{z^2}{\sqrt{3}} \right)
= \frac{a\sqrt{3}}{2} \, z -\frac{a}{2} \, \frac{ \widehat{g}(z) }{z}.
\end{align}
The sum is the transform of
$g(t\geq 4)\,\big[\widehat{g}_{t\geq 4}\big]$, written as
$\widehat{g} - \widehat{g}_{t=2}$.  With Eqs.~(\ref{eq-Hhat}),
(\ref{eq-hhat}), and (\ref{eq-h1hat}), the generating function 
for the amplitude of the process at the root is
\be  \label{eq-HhatResult}
\widehat{H}_n(z) = 
\left[ - \frac{a}{2}  \right]^n \,
\left[ \frac{\widehat{g}(z) - \sqrt{3}\,z^2}{z} \right]^n 
\frac{1}{1-\widehat{g}(z)},
\ee
with $\widehat{g}(z)$ given in Eq.~\eqref{eq-ghat}.
Now we need to invert this.


\subsection{Inverse transform: $H_n(t)$ asymptotic}

We take the inverse $z$--transform via an integral, and using Laurent
expansion and the Residue theorem,
\begin{align*} 
H_n(t) = \ & \frac{1}{2\pi i} \oint_{|z|=r<1}
\frac{ \widehat{H}_n(z) }{ z^{t+1}} dz  \\ 
= \ &  \frac{1}{2\pi i} \left[ - \frac{a}{2}  \right]^n
\oint \frac{1}{z^{t+1}} 
\left[ \frac{\widehat{g}(z) - \sqrt{3}\, z^2}{z} \right]^n
\frac{dz}{1-\widehat{g}(z)}.
\end{align*}
This integral is too complicated to yield a closed-form solution.
We look for its asymptotic behavior in the form
\be \label{eq-stDescForm}
H_n(t) \, = \, \frac{(-a)^n}{2\pi i} \; \frac{1}{2^n}
\oint\limits_{|z|=r} \frac{ \left[
\widehat{g}(z) - \sqrt{3}\,z^2\right]^n }{
1-\widehat{g}(z)} \, \frac{dz}{z^{t+n+1}},
\ee
using the steepest descent method.  The calculation is discussed
in Appendix \ref{app-stDesc}.  The asymptotic of the amplitude of
the process at the root, starting from a level $n$ in the tree,
with $\tau\equiv t-n$, is 
\begin{align} \label{eq-resHn}
H_n(t) \, \sim &  \, \frac{(-a)^n}{2\pi i} \frac{1}{2^n} 
\, \times \, (\sqrt{2})^n  (-1)^n \\
& \notag \times \,
\left[ \, c_{1n} \, e^{ - \bm{i} \gamma n } \,
\frac{ e^{- \bm{i}\,\lambda_1 \frac{ \tau }{2}} }{\tau^{3/2} }
\, - \, c_{2n} \, e^{ \bm{i} \gamma n } \,
\frac{ e^{- \bm{i} \, \lambda_2 \frac{\tau }{2} } }{ \tau^{3/2} }
\, \right].
\end{align}
Constants $c_{1n}$ and $c_{2n}$ are linear in $n$, while $\gamma$
and $\lambda$'s are real constants (Appendix \ref{app-stDesc}).
The probability is
\be \label{eq-stDescProb}
|H_n(t)|^2 \, \sim \, 
\frac{ C^2 - 2\, \mathrm{Re} \left\{ c_1 c_2^{*} \,  e^{ - \bm{i}
\left[  2\gamma n + (\lambda_1 - \lambda_2 )\frac{\tau}{2} 
\right] } \right\} 
}{ 4\pi^2 \; 2^n \; \tau^3 },
\ee
where $C^2=|c_1|^2 + |c_2|^2 \sim n^2$.  The oscillations of the
exponential term are rapid at large times (and/or $n$), and this
function behaves as $\sim \tau^{-3}$. The $n$ dependence is 
$\sim n^2/2^n$ at large times.  Also, we see that the walk is
transient, in the sense introduced in 
\cite{SJK08prl,*SJK08pra,CMGV10}, since
$\sum_{t}^{\infty} |H_n(t)|^2$ is finite.

The observed power law decay differs sharply from the exponential tail
of the classical walk (Appendix \ref{app-cLine}).  On finite graphs
built with binary trees, this exponentially slower decay may lead to
algorithms with significant speedups.  The treatment in this section 
is meant to lay the groundwork for such investigations.

This behavior should also have general implications for physics
of systems modeled with a binary tree.


\subsection{Inverse transform: $H_n(t)$ computed}

\begin{figure}[!ht]
\includegraphics[width=3.3in,clip=true]{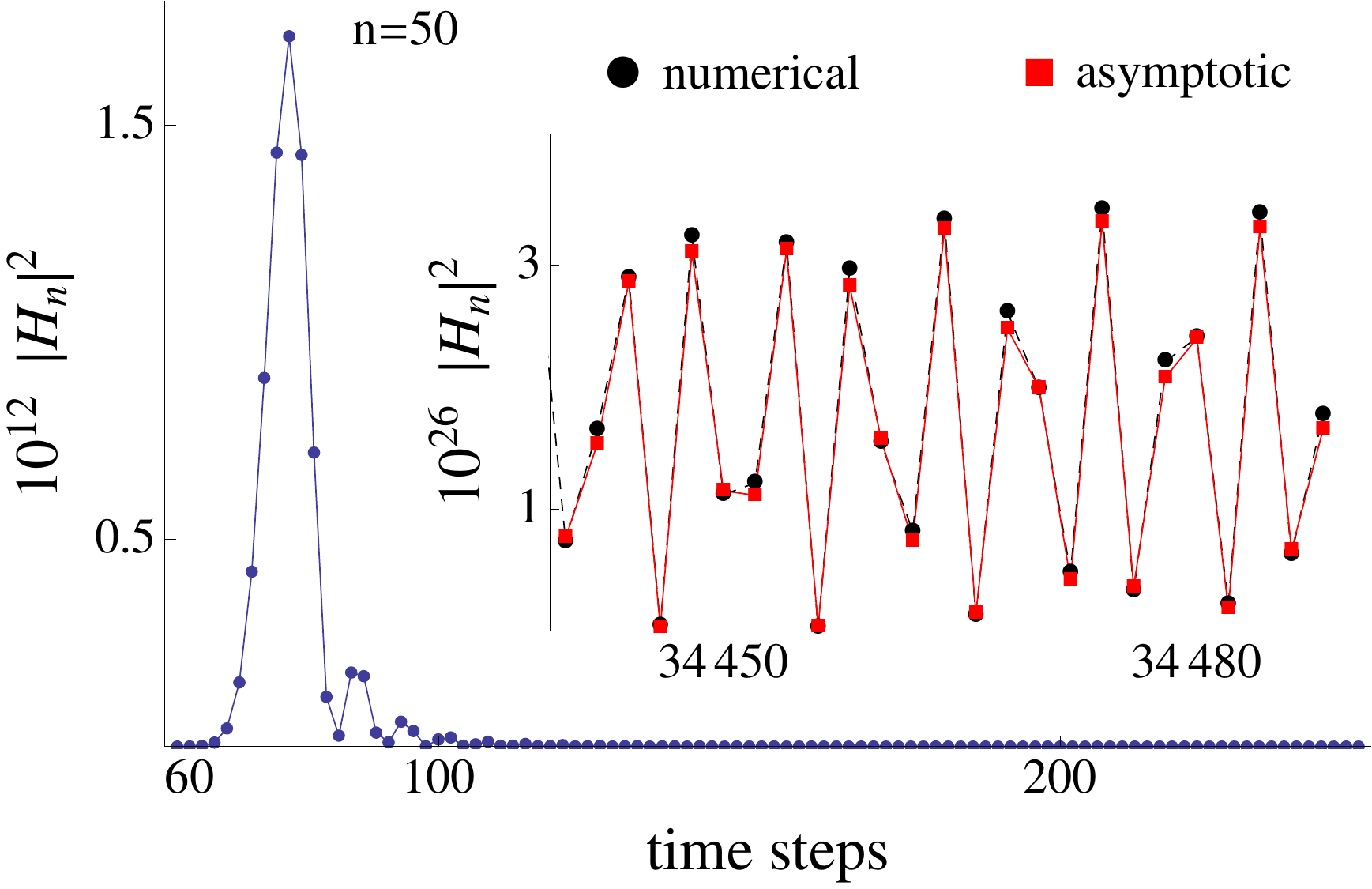}
\caption{ \label{fig-numInset} (Color online)
Probability at the root with time ($n=50$), computed via
Eq.~(\ref{eq-numAmpl}).  Inset compares this (points) with
the steepest descent asymptotic  (\ref{eq-stDescProb})
at same time steps (red squares).
}
\end{figure}

The transform is defined as 
$\widehat{H}_n(z)=\sum_{t\geq n} H_n(t) z^t$,
and using its Taylor expansion and equating coefficients,
\be \label{eq-numAmpl}
H_n(t) = \frac{\widehat{H}_n^{(t)}(z)|_{z=0} }{ t!}.
\ee
This can be evaluated efficiently, for a range of values of $t$
for a fixed $n$, providing the full amplitude. Symbolic calculation of
derivatives (with \textsc{Mathematica}) allows for values of $n$ in the
thousands.  We note that the amplitude shows an interference pattern,
with the main peak followed by (much) smaller, rapidly diminishing,
secondary peaks.  Probability at the root with time is shown in
Fig.~\ref{fig-numInset}, for $n=50$. The shape does not depend on $n$. 
At long times this exact result can be compared with asymptotics
(\ref{eq-stDescProb}), see inset in Fig.~\ref{fig-numInset}.
The tail exhibits $t^{-3}$ dependence, in agreement with
Eq.~(\ref{eq-stDescProb}).

\subsection{Comparison and algorithmic aspects}

\begin{figure}[!ht]
\includegraphics[width=3.2in,clip=true]{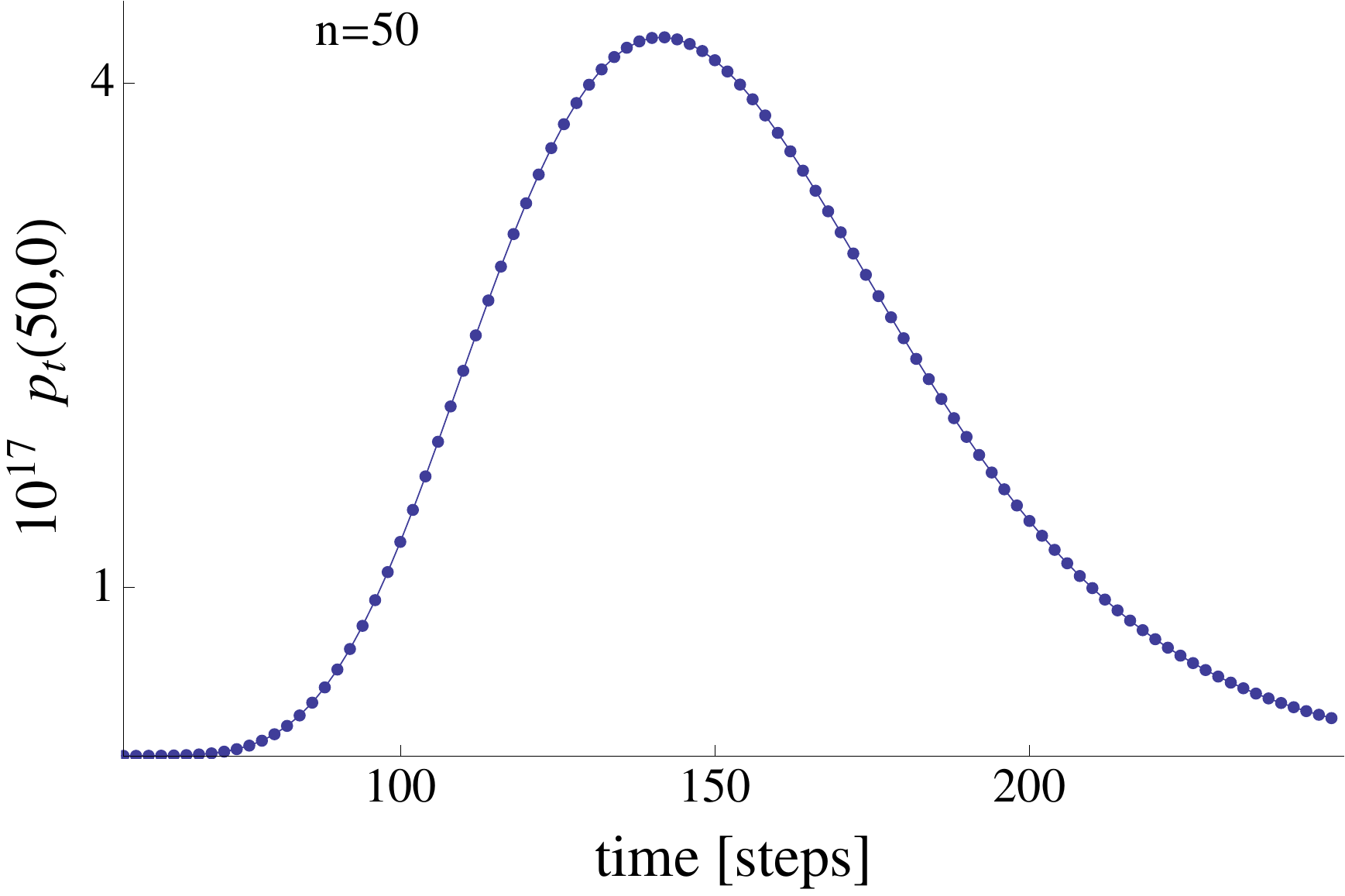}
\caption{  \label{fig-clProb}
(Color online) Classical walk, probability at the root
(Appendix \ref{app-cLine}). 
}
\end{figure}

We considered an algorithm of finding the root starting at the 
$n$--th generation in the tree.  Here we compare the quantum and
classical walks, and their estimated run times, using data calculated
via Eqs.~(\ref{eq-numAmpl}) and (\ref{eqApp-ptReal}).  Probability
peaks and times at which they are reached, for quantum and classical
walks, are shown in Table~\ref{tab-peakComp} for a few initial
levels $n$.
\begin{table*}[!tb]
\begin{ruledtabular}
\begin{tabular}{ | c | c | c | c | c | c | c | } 
initial level $n$ & 10   & 20  
    & 50                         & 100   
    & 200                        & 500    \\ \hline
max $|H_n|^2$ (at  $t$)  & $6.8\times 10^{-4}$ (16)
    & $3.9 \times 10^{-6}$ (30) 
    & $1.7 \times 10^{-12}$ (76) & $5.3 \times 10^{-23}$ (150)
    & $7.6\times 10^{-44}$ (298) & $5.1\times 10^{-106}$ (738) \\
max $p_t(n)$ (at  $t$) & $1.2\times 10^{-4}$ (22)
    & $7.2\times 10^{-8}$ (52) 
    & $4.2\times 10^{-17}$ (142) & $2.6\times 10^{-32}$ (292)
    & $1.4\times 10^{-62}$ (592) & $4.5\times10^{-153}$ (1492) \\
\end{tabular}
\end{ruledtabular}
\caption{  \label{tab-peakComp}
Probability peaks at the root with their times, for quantum
$\big[|H_n(t)|^2\big]$ and classical $\big[p_t(n,0)\big]$ walks,
with $n$.
}
\end{table*}
Probability at the root for the classical walk is shown in
Fig.~\ref{fig-clProb}.

\begin{figure}[!ht]
\includegraphics[width=3.3in,clip=true]{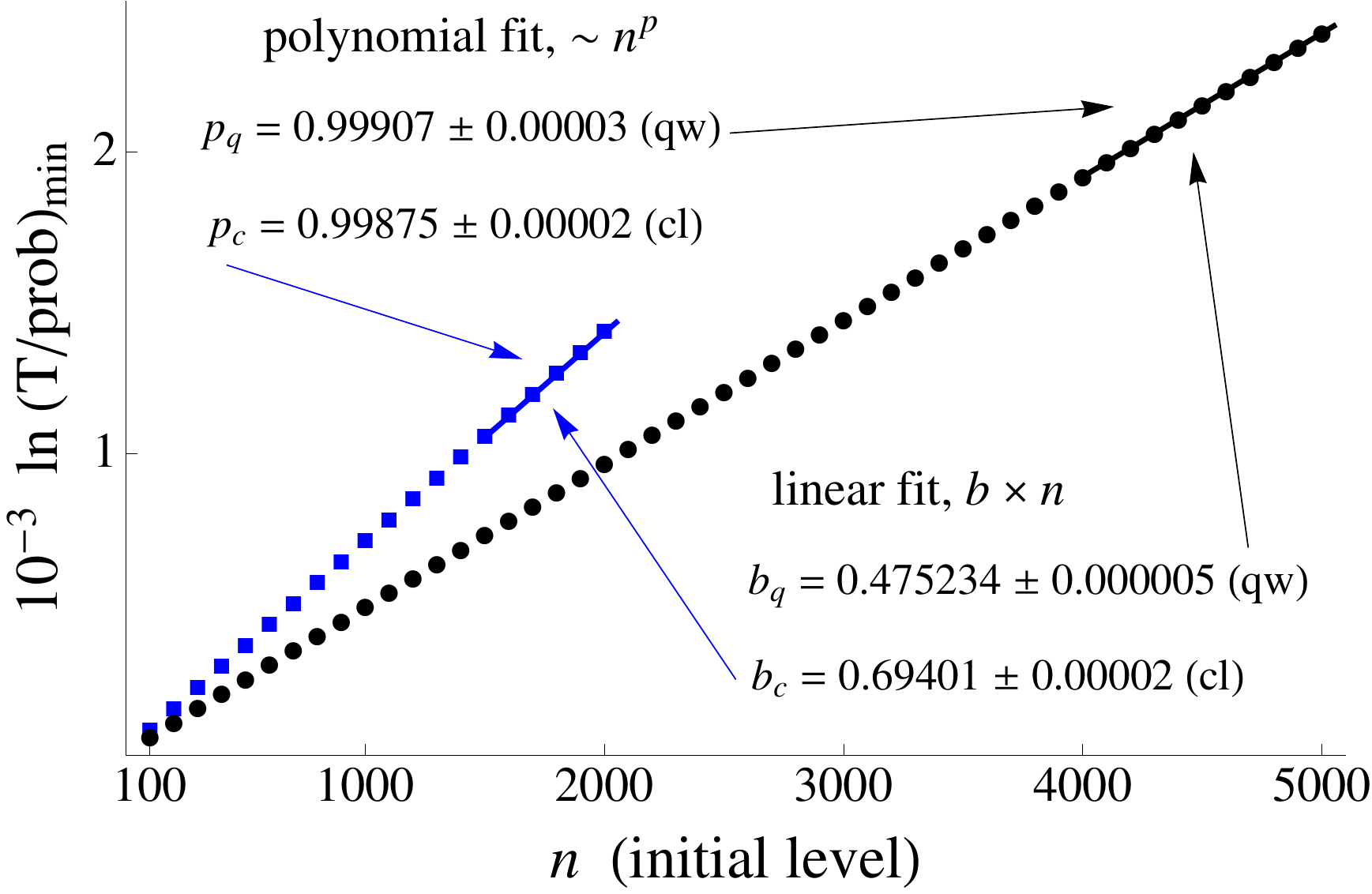}
\caption{  \label{fig-fits}  (Color online)
Natural log ($\ln$) of run time with initial level in the tree, for
quantum (points) and classical (blue squares) walks.  For plot clarity
not all points are shown.  Parameter values for polynomial ($\sim n^p$)
and linear ($bn$) fits are shown, with lines through data used for fits
(see text). The run time is $\sim e^{b n}$, and $b_q/b_c \to 2/3$ 
estimates the quantum over classical walk speedup.
}
\end{figure}

The best run time for a given $n$ is estimated as follows.
The inverse of the probability is the average number of times
needed to run in order to hit the root; multiplying it by its time
gives the total running time to hit the root. We need its minimum
over all time steps, $\min\big\{t\big/|H_n(t)|^2\big\}$, usually
the values for the peak.  We use data up to $n=5000$ in steps of
$100$ for the quantum walk, and up to $n=2000$ in steps of $50$
for the classical walk (numerical integration is more demanding).
Using smaller increments in $n$ does not affect results. We fit
the natural logarithm ($\ln$) of run times with a polynomial and
linear (Fig.~\ref{fig-fits}).

The polynomial fit establishes a linear trend (reached at $n \sim$ a
few hundred), so the run time of the quantum walk is exponential in
the initial distance from the root, $\sim e^{b n}$.  The same holds
for the classical walk, and then the ratio of slopes of their linear
fits compares their run times.

This ratio does not change much over the whole range of data, being
within a few percent of $2/3$. Still, since an exponential complexity
is fully felt at large $n$, the later portions of data are more
relevant for algorithmic comparison.  For the last quarter of
data ranges, indicated in Fig.~\ref{fig-fits} by lines fit through
data, the ratio of quantum to classical slopes is $\approx 0.685$,
within $3\%$ of $2/3$.  Thus it appears reasonable to conjecture
the algorithmic speedup of the order of $2/3$.  (For a run time $T$
for the classical walk, one expects the run time on the order of
$T^{2/3}$ for the quantum walk.)

We note the behavior of peaks times with $n$. For the classical walk,
$t_{\text{max}}^{cl}(n) = 3n - 8$ (exact), while for the quantum walk
$t_{\text{max}}^{qw}(n) \approx 1.46 n$ (where data allows for
a fit of $\sim 1.5n$, and for an $\sim \ln n$ correction).


\section{\label{sec-disc} Concluding Remarks}

Quantum walks are quantum processes with a specific mixing of states;
particular unitary processes.  In this vein, we propose to approach
and study them using ideas from classical random walks with memory.

We have demonstrated how a general framework for discrete-time quantum
walks arises as a natural analog of a specific representation of a
classical memory--$2$ Markov chain.  Walks are implemented by
constructing a local operator with no restrictions other than
unitarity.  The framework needs no ``coin'' degrees of freedom,
is flexible, and is applicable to general graphs.  This approach may
make it easier to obtain walks on structures for which significant
speedups are expected.

The evolution operator works separately on each component of the
amplitude, reducing the state space, and effectively deconstructing
the amplitude [as in in the binary tree example,
Eqs.~(\ref{eq-Uj}--\ref{eq-evolStep})].  This makes quantum-mechanical
correlations and interference transparent in quantum walks, making
their explicit study easier. It should also aid the use of quantum
walks as a general tool for exploration and modeling of physical
systems.

In Sec.~\ref{sec-BT} we use the framework to build a symmetric
discrete-time quantum walk on a semi-infinite binary tree. 
We start the walk at a level $n$ in the tree and find its amplitude
at the root as a function of time and $n$.  The construction of the
walk is simple, but the calculation is complicated by a (reflective)
boundary.  The generating function of the amplitude is found
explicitly, and its asymptotic is found via the steepest descent
method. The full solution is computed numerically.

These solutions show interesting features.  The asymptotic decays
in time by the power law (as opposed to the exponential tail of the
classical walk), representing long--range correlations. This hints
at significant speedups on restricted structures. The amplitude 
exhibits a damped interference pattern, with a distinct and sharp peak.
In comparison with the classical walk, the probability peak is reached
more quickly, and is orders of magnitude greater, already at small
$n$.  The run time for hitting the root on a semi-infinite binary
tree is exponential with $n$ for the quantum walk, as it is for the
classical walk. It is still clearly slower in $n$ and, following
suggestive data trends, we conjecture the polynomial algorithmic
speedup of the order of $2/3$ over a classical walk.


\appendix

\section{\label{app-stDesc} Steepest descent calculation summary}

\begin{figure}[!ht]
\includegraphics[width=3.2in,clip=true]{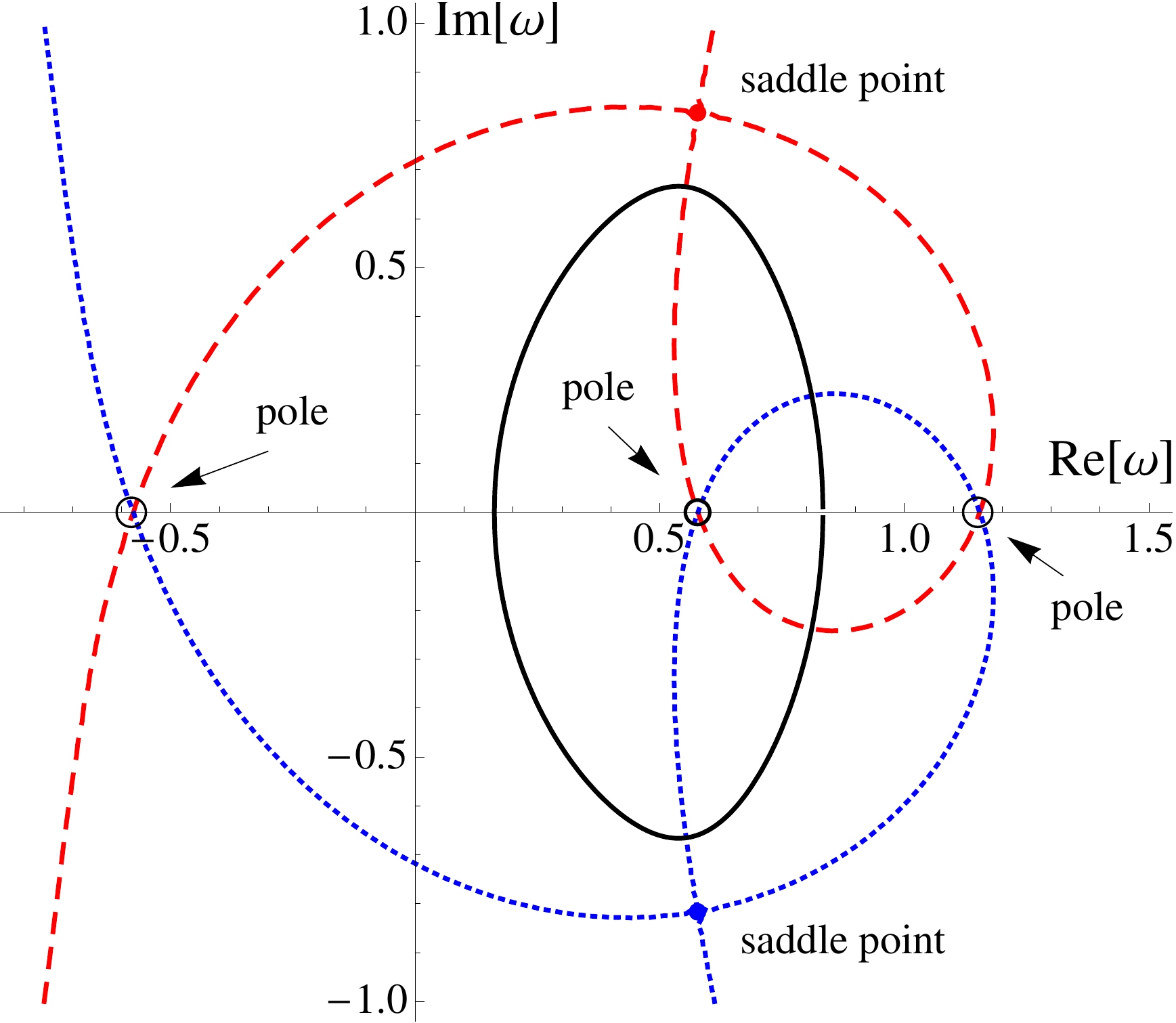}
\caption{ \label{fig-app_sdc}  (Color online)
Original integration contour (solid), steepest descent paths (dashed
and dotted), and critical points.
}
\end{figure}

Integrals suitable for analysis by the steepest descent method are
typically of the form \cite{Wong89,*BlHa86}
\be \label{eqApp_sdm}
I(k) = \int_C f(\omega) \, e^{k\Phi(\omega)} \, d\omega.
\ee
We use Eq.~(\ref{eq-stDescForm}), where the exponent will be formed
from powers of $z$.  As $z$ is always squared we first change
variables via $z^2=\xi$.  Accounting for the double winding,
\ben
I(t; n) = 2 \times\, \frac{1}{2} \oint_{|\sqrt{\xi}|=r} \frac{
\left[ \frac{\widehat{g}(\xi)}{\xi} - { \scriptstyle \sqrt{3} }
\right]^n 
}{ 1 - \widehat{g}(\xi)} \; \frac{1}{\xi} \; 
\frac{d\xi}{\xi^{ \frac{t-n}{2} }}.
\een
We now use $\widehat{g}(\xi)/\xi = \omega$, to transfer some of the
integrand's complexity into the exponent,
\ben
\frac{\widehat{g}(\xi)}{\xi} = \omega, \quad \xi = \varphi(\omega)
= \, a\sqrt{3} \, \frac{\omega - \frac{1}{\sqrt{3}}
}{ (\omega + \frac{1}{\sqrt{3}}) (\omega - \frac{2}{\sqrt{3}}) }.
\een
Carrying out the substitution, we have
\begin{equation} \label{eqApp_intExpand}
I(t;n) = \oint_{|\sqrt{\varphi}|=r} 
\frac{\left( \omega- {\scriptstyle \sqrt{3}} \right)^n
}{\left( 1-\omega\,\varphi \right) } \,
\frac{\varphi\prime }{\varphi} \; \varphi^{ - \frac{t-n}{2} }\,
d\omega,
\end{equation}
in the form (\ref{eqApp_sdm}), with 
$\varphi^{-\frac{t-n}{2}} = e^{ \frac{t-n}{2} \ln(\varphi^{-1})}$,
and
\be \label{eqApp-sdFacts}
f =  \frac{ \left( \omega- {\scriptstyle \sqrt{3}} \right)^n
}{ \left( 1-\omega\,\varphi \right) } \,
\frac{\varphi\prime }{\varphi}, 
\quad \Phi = - \ln\varphi, \quad k= \frac{t-n}{2}.
\ee
Keeping $\varphi\prime/\varphi$ will be useful.  Consider the critical
points. A pole of order $\frac{t-n+2}{2}$ is at
$\varphi=0\Rto\omega_p=1/\sqrt{3}$.  Two simple poles are at
$-\frac{1}{\sqrt{3}},\, \frac{2}{\sqrt{3}}$.  The logarithm's branch
point is at $\varphi = 1$,  and since this is not at 
$\omega_{p}=1/\sqrt{3}$, what the contour must enclose, we can take
any convenient branch.  Two simple saddle points are
\ben
(\ln\varphi)\prime = 0 \ \Rto\ \omega_{s1/s2} = 
\frac{1 \pm \, \bm{i} \, \sqrt{2}}{\sqrt{3}} = 
e^{\pm\, \bm{i}\, \arctan \sqrt{2} }.
\een
The main contribution to this integral comes from saddle points.
A branch of the original integration contour $|\sqrt{\varphi}|=r$
can be chosen (via $r$) for use with steepest descent paths.  There
are no issues with deforming the contour, as no critical points are
in the way, any branch of the logarithm is good, and
$k=\frac{t-n}{2} \in \mathbb{Z}$  \, (Fig.~\ref{fig-app_sdc}).

In the steepest descent method decreasing orders of contribution
are computed mostly via expansion around saddles. (There are theorems
and formulas for the first--order contribution, but here it is zero.)
Around $\omega_s$ we have
$
\Phi(\omega) = \Phi(\omega_s) + \frac{1}{2!}
\big[ \Phi(\omega) \big]_{\omega_s}''
\,(\omega-\omega_s)^2 + o(\omega^2),
$
and the usual change of variables $\Phi(\omega)-\Phi(\omega_s) = -y$
gives 
\be \label{eqApp-omegaY}
\omega = \mp b \sqrt{y} + \omega_s, \quad b = \sqrt{ \frac{2} 
{ \big[ \ln\varphi(\omega) \big]_{\omega=\omega_s}'' } }, 
\ee
where $y$ is zero at the saddle and real along the steepest descent
path. We used 
$\sqrt{ \left( \ln\varphi^{-1} \right)''} 
= \bm{i} \sqrt{ \big( \ln\varphi \big)''}
$.
Now 
$y = \ln\big[ \varphi(\omega) / \varphi(\omega_s) \big]$
and 
$dy = \big[\varphi(\omega)' / \varphi(\omega)\big] \, d\omega$,
and with $\varphi(\omega) = \varphi(\omega_s)\, e^y$ the integral
(\ref{eqApp_intExpand}) and (\ref{eqApp-sdFacts}) becomes
\be \label{eqApp-fInt}
I(k) \sim e^{k\Phi(\omega_s)} \oint \frac{ 
\big[\omega(y) - {\scriptstyle \sqrt{3}} \big]^n \,
}{ 1 - \omega(y) \, \varphi(\omega_s)\,e^y
} \; e^{-ky} \, \,dy.
\ee
Now we can directly expand around $y=0$ (to any order), restricting
integration to the line along the steepest descent path, close to
saddles. The signs in Eq.~(\ref{eqApp-omegaY}) correspond to the
opposite directions from the saddle; we label $+/-$ as ``R/L.''
Substituting $\omega(y)$ and expanding,
\begin{align}
& I \sim A\, \varphi_s^{-k} 
\int_{0}^{\delta} \big( 1 \pm B\,\sqrt{y} \,\big) \; e^{-ky} \, dy,
\quad \delta \sim o(1), \\
& \notag
A = \frac{ \big[ \omega_s - {\scriptstyle \sqrt{3}} \big]^n
}{ 1-\omega_s\, \varphi_s }, \ B = \left(
\frac{ b\, \varphi_s }{ 1 - \omega_s \varphi_s }
+ \frac{b}{ \omega_s - {\scriptstyle \sqrt{3}} } \; n
\right).
\end{align}
For compactness we use
$\varphi_s \equiv \varphi(\omega_s) = e^{\bm{i} \lambda_s}$, with
$\lambda_{s1} = \arctan \left[
({\scriptstyle 9\sqrt{3}+8\sqrt{2})/23} \right]$,
$\lambda_{s2} = \arctan \left[
({\scriptstyle 9\sqrt{3}-8\sqrt{2})/23} \right]-\pi$.
Note that $A(n) \sim (\sqrt{2})^n$, as 
$\omega_s-{\sqrt{3}} = \sqrt{2}e^{\bm{i}\gamma_s}$, with
$\gamma_{s1/s2} = \mp \left[
\arctan(1/{\scriptstyle \sqrt{2}}) - \pi\right]
\equiv \mp (\gamma - \pi)$, and we will extract $\pi$ later.
The integral is dominated around $\omega_s$ ($\delta\approx 0$),
so it can be formally extended, $\delta\to\infty$, and we get
$I\sim A \int_0^\infty \left(1\pm  B\sqrt{y}\right)\, e^{-ky}\, dy$.
This results in
\be
I_{R/L} \, (k;n) \, \sim \, A_n \, \left( 
\frac{1}{k} \pm \frac{\sqrt{\pi}}{2}\, \frac{B_n}{k\sqrt{k}}
\right) \, \varphi(\omega_s)^{-k}.
\ee
Subtracting contributions along opposite directions, the first 
non--zero order for either saddle is
\be
I_s \sim (\sqrt{2})^n \, e^{\bm{i}\,\gamma_s n} \; 
(a_{s} + d_{s}\,n) \; 
\frac{ e^{-\bm{i}\, \lambda_{s}  k} }{ k\sqrt{k} }.
\ee
We broke up the $A_nB_n$ term found in $I_R-I_L$, to show the
structure of $n$ dependence, where 
$a_s = \frac{ b\,\varphi_s \sqrt{\pi} }{ (1-\omega_s\varphi_s)^2 }$
and $d_s = \frac{ b\,\sqrt{\pi} }
{ (1-\omega_s\varphi_s)(\omega_s - {\scriptstyle \sqrt{3}}) }$ are
of order $\sim 1$. Here we extract $\pi$ from $\gamma_s$, and will
use $e^{\bm{i}\gamma_s} = (-1) e^{\mp\bm{i}\gamma}$.  Contributions
for saddles are subtracted (for consistency of $\pm \sqrt{y}$
directions) and, with 
$c_{s,n} \equiv a_s+d_s n$, we get Eq.~(\ref{eq-resHn}).

The full expansion of the integral (\ref{eqApp-fInt}) results in
nested sums of a Gamma function. This cannot capture the peak of the
amplitude though, and is not needed for our asymptotic analysis,
so we do not pursue it here.

\section{\label{app-cLine} Random walk on a semi-infinite line}

For completeness here we provide the application of the method developed
in \cite{Kovc09,Kovc10} (based on Karlin-McGregor spectral approach to
random walks) to a classical walk on a binary tree.

If $P$ is a reversible Markov chain over a sample space $\Omega$,
and $\pi$ is a reversibility function (not necessarily a probability
distribution), then $P$ is a self-adjoint operator in $\ell^2(\pi)$,
the space generated by the inner product
$$<f,g>_{\pi}=\sum_{x \in S} f(x)g(x)\pi(x)$$
induced by $\pi$. If $P$ is tridiagonal operator (i.e. a nearest-neighbor
random walk) on  $\Omega=\{0,1,2,\dots\}$, then it must have a simple 
spectrum, and is diagonalizable via orthogonal polynomials, as it was 
studied in the 1950s and 1960s by Karlin and McGregor.  There the extended
eigenfunctions $Q_j(\lambda)$ ($Q_0 \equiv 1$) are orthogonal polynomials
with respect to a probability measure $\psi$ and
$$p_t(i,j)=\pi_j \int_{-1}^1 \lambda^t Q_i(\lambda) Q_j(\lambda)
d\psi(\lambda) \quad \forall i,j \in \Omega, $$
where $\pi_j$ ($\pi_0=1$) is the reversibility measure of $P$.
Consider the following Markov chain 
$$P=
\begin{pmatrix}
0 & 1 & 0 & 0 & \dots \\
q & 0 & p & 0 & \dots \\
0 & q & 0 & p & \ddots \\
\vdots & \vdots & \ddots & \ddots & \ddots
\end{pmatrix}
\qquad p>q.$$
Orthogonal polynomials are obtained via solving a simple linear
recursion: $Q_0=1$, $Q_1=\lambda$, and
\ben 
Q_n(\lambda)=c_1(\lambda) \rho^n_1(\lambda)
+ c_2(\lambda)\rho^n_2(\lambda),
\een 
where  $\rho_1(\lambda)= \frac{\lambda+\sqrt{\lambda^2-4pq} }{ 2p}$
and $\rho_2(\lambda)= \frac{\lambda-\sqrt{\lambda^2-4pq} }{ 2p}$
are the roots of the characteristic equation for the recursion,
and $c_1=\frac{\rho_2-\lambda }{ \rho_2 -\rho_1}$ and 
$c_2=\frac{\lambda-\rho_1 }{ \rho_2 -\rho_1}$.  Now $\pi_0=1$ and
\newline
$\pi_n=\frac{p^{n-1} }{ q^n}$ ($n\geq 1$).  Also, we observe that 
\begin{align*}
& |\rho_2(\lambda)| > \sqrt{ q/p } 
    \quad \text{ on }[-1,-2\sqrt{pq}), \\
& |\rho_2(\lambda)| < \sqrt{ q/p } 
    \quad \text{ on }(2\sqrt{pq}, 1], \\
& |\rho_2(\lambda)| = \sqrt{ q/p } 
    \quad \text{ on }[-2\sqrt{pq}, 2\sqrt{pq}],
\end{align*}
and $\rho_1\rho_2=\frac{q }{ p}$. The above will help us to identify
the point mass locations in the measure $\psi$ since each point mass
in $\psi$ occurs when $\sum_{k} \pi_k Q_k^2(\lambda) < \infty$. 
Thus we need to find all $\lambda \in (2\sqrt{pq}, 1]$ such that
$c_1(\lambda)=0$ and all $\lambda \in [-1,-2\sqrt{pq})$ such that 
$c_2(\lambda)=0$. But there are no such roots, as $c_1(-1)=0$ and 
$c_2(1)=0$, while $-1 \not\in (2\sqrt{pq}, 1]$ and 
$1 \not\in [-1,-2\sqrt{pq})$. 
Thus there are no point mass atoms in $\psi$, and the mass of $\psi$
must be continuously distributed inside $[-2\sqrt{pq}, 2\sqrt{pq}]$. 
In order to find the density of $\psi$ inside
$[-2\sqrt{pq}, 2\sqrt{pq}]$ we need to find $[e_0,(P-sI)^{-1}e_0]$
for $\operatorname{Im}(s) \not=0$, i.e. the upper left element in
the resolvent of $P$.  

Let  $(a_0(s),a_1(s),\dots)^T=(P-sI)^{-1}e_0$, then
$$-sa_0+a_1=1,\quad\text{and}\quad qa_{n-1}-sa_n+pa_{n+1}=0$$
Thus $a_n(s)=\alpha_1 \rho_1(s)^n+\alpha_2 \rho_2(s)^n,$
with $\alpha_1= \frac{a_0(\rho_2-s) -1 }{ \rho_2(s)-\rho_1(s)}$
and $\alpha_2= \frac{1-a_0(\rho_1-s) }{ \rho_2(s)-\rho_1(s)}$.
Since 
$(a_0,a_1,\dots) \in \ell^2(\mathbb{C}, \pi)$,
$$|a_n| \sqrt{ \frac{p^n}{q^n} } \rightarrow 0 \qquad \text{as} 
\quad n \rightarrow +\infty$$
Hence when $|\rho_1(s)| \not= |\rho_2(s)|$,
either $\alpha_1=0$ or $\alpha_2=0$, and therefore
\begin{equation}\label{a0}
a_0(s)= 
\frac{ \mathbbm{1}_{|\rho_1(s)|< \sqrt{ \frac{q}{p} }} }{ \rho_1(s)-s }
+ \frac{ \mathbbm{1}_{|\rho_2(s)|<\sqrt{ \frac{q}{p}} } }{ \rho_2(s)-s }.
\end{equation}
Also $d\psi(z)=\varphi(z)dz$, where $\varphi(z)$ is an atom-less density
function over $[-2\sqrt{pq}, 2\sqrt{pq}]$, and 
$$a_0(s)=\int_{-2\sqrt{pq}}^{+2\sqrt{pq}} 
\frac{ d\psi(z) }{ z-s }
=\int_{-2\sqrt{pq}}^{+2\sqrt{pq}} \frac{\varphi(z)dz }{ z-s}.$$
Next we use the following basic property of Cauchy transforms 
$Cf(s)=\frac{1 }{ 2\pi i} \int_{\mathbb{R}} \frac{f(z)dz }{ z-s}$ that  
can be derived using the Cauchy integral formula, or similarly, 
an approximation to the identity formula:
\begin{equation} \label{cauchy}
C_+-C_-=I.
\end{equation} 
Observe that the curve in the integral need not be in $\mathbb{R}$
for $C_+-C_-=I$ to hold.  Here
\begin{align*}
C_+f(z) & = \,
\lim_{s \rightarrow z:~ \operatorname{Im}(s)>0} Cf(s),
\quad\text{and} \\ 
C_-f(z) & = \,
\lim_{s \rightarrow z:~ \operatorname{Im}(s)<0} Cf(s),
\end{align*}
for all $z \in \mathbb{R}$.
The relation (\ref{cauchy}) implies
\ben
\varphi(x)=\frac{ 1}{ 2\pi i}
\left(\lim_{\substack{s=x+i\varepsilon\colon \\ \varepsilon 
\rightarrow 0+}}  a_0(s) - 
\lim_{\substack{s=x-i\varepsilon \colon \\ \varepsilon \rightarrow 0+}}
a_0(s) \right),
\een
for all $x \in (-2\sqrt{pq}, 2\sqrt{pq})$.  Recalling (\ref{a0}),
we express $\varphi$ as
$\varphi(x)= 
\frac{\rho_1(x)-\rho_2(x) }{  2\pi i(\rho_1(x)-x)(\rho_2(x)-x)}$
 for  $x \in (-2\sqrt{pq}, 2\sqrt{pq})$, which in turn simplifies to
$$\varphi(x)=\begin{cases} 
  \frac{\sqrt{4pq-x^2} }{ 2\pi q(1-x^2)}  & \text{ if } 
  x \in (-2\sqrt{pq}, 2\sqrt{pq}), \\
  0 & \text{ otherwise. }
\end{cases}$$
Here $\varphi(x)$ always integrates to $1$ over
$(-2\sqrt{pq}, 2\sqrt{pq})$.  Now
\begin{align*}
p_t(n,0) & = \int_{-2\sqrt{pq}}^{+2\sqrt{pq}}
\lambda^t Q_n(\lambda)\varphi(\lambda) d\lambda \\
& = \int_{-2\sqrt{pq}}^{+2\sqrt{pq}} \lambda^t 
(c_1\rho_1^n+c_2\rho_2^n) 
\frac{ (\rho_1 -\rho_2 ) \, d\lambda }{
2\pi i (\rho_1- \lambda)(\rho_2-\lambda)},
\end{align*}
and therefore, since
$c_1= \frac{\rho_2-\lambda }{ \rho_2 -\rho_1}$ and
$c_2= \frac{\lambda-\rho_1 }{ \rho_2 -\rho_1}$,
\be \label{eqApp-ptReal}
p_t(n,0)= \frac{1 }{ 2\pi i}
\int_{-2\sqrt{pq}}^{+2\sqrt{pq}} \lambda^t
\left( \frac{\rho_2^n}{\rho_2-\lambda}
- \frac{\rho_1^n }{ \rho_1-\lambda}\right)d\lambda.
\ee
This can be treated as a complex integral, for example, with steepest
descent.  But one can observe directly in Eq.~(\ref{eqApp-ptReal})
that the tail of $p_t(n,0)$ decays as $(2\sqrt{pq})^t$ when
$t \rightarrow +\infty$.  Thus, using $p=2/3$ and $q=1/3$ for the
classical symmetric walk on the semi-infinite binary tree, the decay
rate will be $(2\sqrt{2}/3)^t$, giving us the exponential asymptotics.
The probability integral (\ref{eqApp-ptReal}) can be efficiently
evaluated numerically (see Fig.~\ref{fig-clProb}).


\bibliography{bibBinTree}


%

\end{document}